\documentclass[journal,comsoc]{IEEEtran} 

\usepackage[T1]{fontenc}

\usepackage{amsmath}
\usepackage{amsthm} 
\usepackage[cmintegrals]{newtxmath}
\usepackage{cite}
\usepackage[utf8]{inputenc}
\usepackage[english]{babel}
\usepackage[]{verbatim}
\usepackage{syntonly}
\usepackage{textcomp}
\usepackage{amsmath}
\usepackage{ upgreek }
\usepackage{graphicx} 
\usepackage{varioref}
\usepackage{hyperref}
\usepackage[noabbrev]{cleveref}
\usepackage{textcomp,xcolor}

\newtheorem{Corollary }{Corollary}

\begin{document}
	
	
	\title{Multi-Mode High Altitude Platform Stations (HAPS) for Next Generation Wireless Networks}

	\author{Safwan Alfattani, Wael Jaafar, Halim Yanikomeroglu, and Abbas Yongaçoglu
 \thanks{
 This work is funded by a scholarship from King Abdulaziz University, Saudi Arabia and the Natural Science and Engineering Research Council of Canada (NSERC).}
	}

	\maketitle
	
	\begin{abstract}
The high altitude platform station (HAPS) concept 
 has recently received   notable attention from both industry and academia to support future wireless networks. 
A HAPS can be equipped with 5$^{\rm th}$ generation (5G) and beyond technologies such as massive multiple-input multiple-output (MIMO) and reconfigurable intelligent surface (RIS). Hence, it is expected that HAPS will support numerous applications in both rural and urban areas. However, this comes at the expense of  high energy consumption and thus shorter loitering time. 
 To tackle this issue, we envision the use of a multi-mode HAPS that can adaptively switch between different 
 modes so as to reduce energy consumption and extend the HAPS loitering time. These modes comprise a HAPS super macro base station (HAPS-SMBS) mode for enhanced computing, caching, and communication services, a HAPS relay station (HAPS-RS) mode for active communication, and a HAPS-RIS mode for passive communication.
 This multi-mode HAPS ensures that operations rely mostly on the passive communication payload, while switching to an energy-greedy active mode only when necessary. In this article, we begin with a brief review of HAPS features compared to other 
 non-terrestrial systems, followed by an exposition of the different HAPS modes proposed.
 Subsequently, we illustrate the envisioned multi-mode HAPS, and discuss its benefits and challenges. Finally, we validate the multi-mode efficiency through a case study.

	\end{abstract}
	
	
\vspace{-10pt}	
	
	\section{Introduction}
Beyond fifth-generation (B5G) and sixth-generation (6G) technologies are expected to support novel use cases thanks to their anticipated ubiquitous and reliable connectivity with a massive numbers of devices with high data rates and low latency. 
But it has been shown to be unfeasible and cost-inefficient to attempt to fulfill the requirements of future networks by relying solely on terrestrial networks.
To complement terrestrial systems, non-terrestrial networks (NTN) are envisioned as a key enabler for next-generation networks. More specifically, given the intrinsic features of NTN, including flexible deployment, strong channel links, and wide coverage footprints, the latter can support terrestrial 
networks by enhancing aspects such as communication, computing, and caching capabilities.

Typically, three types of 
NTN systems are proposed to support future networks, namely unmanned aerial vehicles (UAVs) \cite{Geraci2022}, high altitude
platform station (HAPS\footnote{In
line with the convention in recent ITU (International
Telecommunications Union) documents, in this paper the abbreviation
HAPS is used to denote both the singular and plural usage.}) systems \cite{kurt2021vision}, and low-Earth-orbit (LEO) satellites \cite{Wang2022}. These NTN systems have different features such as operating altitude, size, payload, flight duration, and communication capabilities. 
While UAVs are limited by energy consumption and flight time, LEO satellites suffer from significant path-loss, high mobility, and long communication delays. By contrast, HAPS systems have the largest platform size and payload, less path-loss and delay than LEO satellites, and they can sustain longer missions than UAVs. Accordingly, the development of HAPS technologies has attracted significant interest from both academia  and the industry {\cite{kurt2021vision}}. 
Examples of current HAPS projects include X-station by StratXX, Zephyr by Airbus, Stratobus by Thales, Hawk30 by HAPSMobile, and Phasa-35 by BAE Systems.

One of the 
main issues in HAPS research is the design of the
 communication payload subsystem, as it impacts the range of supported applications, energy consumption, flight duration, and deployment costs. Traditionally, HAPS were developed to serve rural and hard-to-reach areas, 
 and the communication payload was designed in a single mode
 to operate either as a base station (HAPS-BS) or as a relay station (HAPS-RS). 
 An advanced HAPS-BS, referred to as HAPS super macro base station (HAPS-SMBS), was
 recently proposed in \cite{alam2021high}, 
 where it was used
 in urban areas for novel applications beyond connectivity, such as computing, storage,  and sensing. 
 Similarly, an energy-efficient version 
 of a HAPS-RS was recently proposed in {\cite{alfattani2021aerial}}, where a HAPS is equipped with a reconfigurable intelligent surface (RIS), aiming to provide relaying functions. The latter is called HAPS-RIS.
 
Nevertheless, designing a HAPS with a single payload mode either increases its energy consumption or limits its communication capabilities. Hence, to cope with the user traffic and service demand dynamics in the most efficient and cost-effective manner, we propose here the design of a multi-mode HAPS payload, where the HAPS can adaptively switch between different operating  
modes, i.e., HAPS-SMBS, HAPS-RS, and HAPS-RIS, based
on the received demands. Consequently, the usage of active components on the HAPS will be minimized and a more energy-efficient operation will thus be achieved. 

\vspace{-10pt}		
\section{HAPS versus other NTN Systems}

{We discuss here the characteristics of HAPS and its difference from other NTN systems such as UAVs and LEO satellites, in terms of communication
payload capabilities, operations, and suitable applications.}

    
\subsection{HAPS versus LEO Satellites}
HAPS systems have  unique properties. First, HAPS are typically located at an altitude of 20 km, against altitudes between 350 and 2000 km for LEO satellites. 
Hence, a HAPS would experience less path-loss and stronger line-of-sight (LoS) links. The high quality of HAPS communication links to the ground allows it to connect directly to the user equipment (UE) without requiring a special receiver design. This is in contrast to current LEO systems, where sophisticated receivers with high antenna gain are required{\footnote{Recently, satellite direct-to-device solutions are being tested and validated with standard UE. However, until today, the only validated services are limited to emergency messaging and localization.}}.
Also, due to its low altitude, a HAPS is more appealing for delay-sensitive and
critical applications than a LEO satellite.
Second, HAPS systems are quasi-stationary either through fixed-wing HAPS circular trajectories or airships loitering, whereas LEO satellites orbit the Earth at high speeds. Thus, unlike HAPS, satellites suffer from significant Doppler effects, frequent handover, and wasted capacity, due to orbiting under-populated areas. Moreover, given the continuous movements of LEO satellites, a tracking system in current receivers is required.
 Third, HAPS are giant platforms, e.g., aerostatic HAPS have lengths between 100 and 200 m, and aerodynamic HAPS have wingspans between 35 and 80 m. This is up to 20 times the size of a standard LEO satellite. Such HAPS sizes allow to accommodate several communication technologies, including massive MIMO and large RIS. Moreover, HAPS can host heavier payloads, e.g., storage and computing equipment.
Finally, the lifetime of HAPS is estimated to be between a few months and several years, depending on the nature of its mission. Although this is lower than the LEO satellite 10-year life, HAPS will be recoverable at the end of its lifetime.
Moreover, HAPS can be maintained either in the sky or by bringing it back to Earth, which makes it possible to extend its lifetime. 

\vspace{-10pt}	
\subsection{HAPS versus  UAVs}

UAVs can achieve reliable and low-latency communications with ground UEs over small distances of up to a few hundred meters. 
In contrast, HAPS enjoys a wider footprint radius ranging between 40 and 100 km for high throughput communications, and this can be extended to 500 km (ITU-R F.1500). To achieve an equivalent footprint to HAPS, the deployment of costly UAV swarms is needed. 
Also, a better LoS probability can  be realized with HAPS, while UAV links are sensitive to
blockages and high-rise buildings. Given that HAPS can be powered by renewable energy sources, e.g., solar panels, or hydrocarbon fuel (backed up by batteries and fuel cells) \cite{kurt2021vision}, they can sustain longer missions than the battery/hydrocarbon fuel-limited UAVs\footnote{Note that the consumed energy by the communication payload is significantly lower than that required by the flying system.}. 
Finally, the small size of UAVs limits their communication payload and potential applications. For instance, authors of \cite{alfattani2021link} showed that RIS-equipped UAVs perform worse than RIS-equipped HAPS. For the same reason, high storage and computation power cannot be deployed on UAVs, in contrast to HAPS.

	\section{Single-Mode HAPS Communication Payload}
	A HAPS consists of three onboard subsystems: an energy management subsystem, a flight subsystem, and a communication payload subsystem \cite{kurt2021vision}. The energy management subsystem is responsible for power generation using photovoltaic (PV) panels and/or hydrocarbon fuel and for energy storage through Lithium-ion batteries or fuel cells. Moreover, this subsystem controls the energy consumption required by the other subsystems. The flight subsystem controls the mobility and stabilization of the HAPS, whereas the communication payload subsystem mainly manages the communications between the HAPS and other aerial or terrestrial nodes, while also processing and storing other required data. Based on the capabilities of the HAPS in terms of communication, computing, and storage, its power requirements and applications may vary. Typically, three types of HAPS communication payload have been defined, namely  HAPS-SMBS \cite{alam2021high}, HAPS-RS, and HAPS-RIS \cite{alfattani2021aerial}.
	The type of communication payload impacts the potential applications supported, onboard consumed energy, and thus the flight duration of the HAPS. In what follows, we discuss the properties and potential use cases of each HAPS-equipped communication payload type.
	
	\subsection{HAPS-SMBS}
	The main role of the HAPS-SMBS communication payload involves radio frequency (RF) filtering, frequency conversion, and signal amplification. Its 
	multiple antennas transceivers can also encode/decode, precode, and modulate/demodulate signals, as well as switch and route data. The communication payload of the HAPS-SMBS used exclusively for communications is called a ``\textit{regenerative payload}''  by the 3rd Generation Partnership Project (3GPP) standards (TR 38.811), and it supports
	similar tasks to a ground base station or a Node B (gNB). 
    A HAPS-SMBS's ``\textit{regenerative payload}'' can fully process signals and serve users directly, unlike  other communication payload types.
    When the communication payload of a HAPS-SMBS integrates computation and caching capabilities, its role can be extended 
    beyond simple data transmission and reception for users in rural
    and underserved areas. Indeed, HAPS-SMBS can work in tandem with terrestrial networks in 
    dense urban areas to provide 
    numerous applications and novel services  
for 5G and
beyond networks.
    We discuss some  unique HAPS-SMBS use cases below.    
 
	\subsubsection{Increasing network capacity} 
	To cope with the  increasing demands in metropolitan areas, network operators have traditionally relied on densification with macro and small gNBs. However, this might not be a cost-effective solution in dynamic and highly mobile environments. In addition, small-cell densification might not be sufficient to absorb the ever-increasing traffic of connected devices. 
	Moreover, the fixed deployment of terrestrial networks 
	is unable to handle unpredictable congestion
	caused by temporary events. To tackle this issue, a HAPS-SMBS can complement a terrestrial network by providing wide coverage, continuous, and agile connectivity to the terrestrial network's cell-edge and high-traffic demanding UEs.

\subsubsection{Supporting aerial networks and aerial users}
The deployment of UAVs as BSs or relays is seen as a key enabler of future networks, 
given the flexibility and mobility of UAVs.
However, their processing and computation powers are limited, and so they may not be able to execute all required tasks. Hence, UAV-BSs can rely on a HAPS-SMBS to offload their tasks and serve as a reliable computing/storage server. On the other hand, the massive deployment of UAV users (UAV-UEs) is envisioned in the near future for numerous applications such as mail/package delivery, search and rescue, and traffic monitoring \cite{Cherif2021}. However, supporting and managing UAV-UEs through current gNBs might not be practical, since they are designed to serve ground UEs. 
But a HAPS-SMBS would be able to manage UAV-UE swarms
and support their command and control functions.
	\subsubsection{Supporting intelligent transportation systems}
	Intelligent transportation system (ITS) is a paradigm shift that 
	encompasses technologies such as Internet-of-things (IoT) and connected autonomous vehicles (CAVs). This shift aims to enhance the safety, mobility, productivity, and comfort of transportation.
	Efficient ITS services require the implementation of several tasks, such as traffic and road condition monitoring, accident reporting, and vehicle platooning, which depends on the availability of connectivity, computing, and storage capabilities along roads and highways.
    In this context, given a HAPS-SMBS's mobility, sustainability, and communication payload capacity, it is envisioned as a key enabler  of future ITS services, particularly on highways \cite{jaafar2022haps}. 
	
	\subsubsection{Providing aerial datacenter services} Given the high payload of 
	 a HAPS-SMBS, which is up to
	a few hundreds of kilograms \cite{kurt2021vision},
    it can host datacenter equipment to support intensive computation applications, such as virtual/augmented reality (VR/AR). Hence, IoT and low-power devices can offload their computation tasks to the HAPS-SMBS. Also, in case of a terrestrial infrastructure failure, 
    the HAPS-SMBS can provide backup
	storage and computing services.

 \vspace{-10pt}	
	\subsection{HAPS-RS}
	The HAPS-RS can be seen as a lighter version of the HAPS-SMBS. As such, it focuses only on communications and thus consumes less power than the HAPS-SMBS. 
	This is mainly due to the absence of the energy-greedy computing and caching functionalities.  
 HAPS-RS's 
 payload is referred to as a
 ``bent pipe payload'' (TR 38.811), whose functions are limited to RF filtering, frequency conversion, and signal amplification. Typically, a HAPS-RS ensures communications through a ground gateway. Hence, any communication to/from a ground user has to go through two links, namely the gateway-HAPS-RS link and the user-HAPS-RS link. Compared to the previous mode, the propagation delay in the HAPS-RS mode is longer.
 Accordingly, a HAPS-RS supports lighter applications than a HAPS-SMBS, such as:
	\subsubsection{Backhauling traffic from small cells}
	 Wired backhauling infrastructure might not be cost-effective for small cells, especially in rural or hard-to-reach areas due to limited population sizes and the high cost of connecting such areas to the core network through fiber links.
In such cases, a HAPS-RS can ensure the connectivity of small BSs to the core network using RF or free-space optical (FSO) links for higher backhaul capacity. 

 
	\subsubsection{Supporting IoT networks}
	IoT and massive machine type communications have been promoted  as key enablers of future industrial networks. IoT devices are expected to be widely deployed even in areas with limited cellular coverage, such as highways, forests, and oceans. For these particular environments, connectivity through satellites has been previously adopted. However, relying on satellites may be suitable only for delay-tolerant applications, while services with stringent quality-of-service (QoS) demands would require a more effective solution.
	To satisfy such demands, a HAPS-RS can be leveraged for IoT data collection and forwarding to the ground control center through its gateway.
	
 
	\subsubsection{Establishing and maintaining inter-HAPS links}
	To support  connectivity over wide areas, such as countries and trans-continental highways, 
	several HAPS-RS nodes can be deployed. Together, the nodes can form a HAPS constellation with several gateways and inter-HAPS links to enable mesh networking among them and bypass any failure within the constellation. 
	These inter-HAPS links can be achieved through the HAPS-RS mode using either RF or FSO signals \cite{jaafar2022haps}.
	

 \vspace{-10pt}	
	\subsection{HAPS-RIS}
	Both HAPS-SMBS and HAPS-RS require active communication payloads, which consume considerable amounts of energy. Although a HAPS-RS consumes less energy than a HAPS-SMBS, the energy consumed by the former nevertheless impacts the operating lifetime of the HAPS.
	To further reduce power consumption and extend the loitering time of  HAPS, a passive communication payload can be utilized. 
	Passive communication payloads have already been adopted in satellites, such as the Echo satellite, the Lincoln calibration sphere, and the Passive Geodetic Earth Orbiting Satellite (PAGEOS), launched in 1960, 1965, and 1966, respectively, and all using passive reflective satellite surfaces.
	Nevertheless, due to the high altitude of these satellites, applications were limited to low-rate and delay-tolerant ones.
	
	Recently, RIS has attracted interest as a novel passive communication technology.  More specifically, RIS is a nearly passive technology that uses massive reflective elements to alter and reflect the incident RF signals through phase shifting. Since RIS phase shifting can be smartly and dynamically adjusted, the RIS performs better than conventional passive reflection \cite{alfattani2021aerial}. Also, due to the absence of RF chains in RIS,
	it is more energy-efficient than conventional relays.

    Given the opportunities offered by RIS, its integration into HAPS as a light communication payload has already attracted our interest \cite{alfattani2021link}.
 A comparison between HAPS-RS and HAPS-RIS has shown that HAPS-RIS is more energy-efficient and cost-effective. Moreover, most of the applications targeted  by HAPS-RS, such as backhauling, supporting IoT systems, and inter-HAPS links, can be handled by HAPS-RIS. Other use cases can be also listed, including the following:
	\subsubsection{Securing communications}
	Due to its unique properties for signal manipulation, including reflection and absorption, RIS can be used to enhance physical layer security (PLS). This was studied in UAV systems \cite{long2020reflections}, where a UAV-RIS trajectory was jointly optimized with phase shifting
 to maximize the secrecy energy efficiency. The results demonstrated the UAV-RIS's ability to enhance secure energy efficiency by up to 38\% compared to a benchmark without an RIS.
    With  HAPS-RIS, PLS can be extended to large areas where legitimate signals are focused towards their users, while nulling the signals' energy in an eavesdropper's direction. Also, HAPS-RIS can send jamming signals to an eavesdropper to prevent  decoding legitimate signals. Finally, HAPS-RIS can block the jamming signal of an attacker through signal absorption. 
    
	
	\subsubsection{Supporting beyond-cell communications}    
	In dense urban environments, communication links between gNBs and users may be degraded because of 
	blockages or overloaded gNBs.
	To bypass these issues and connect dropped users by the gNBs, HAPS-RIS can be leveraged to connect those users through a dedicated ground station \cite{alfattani2022beyond}. 

	
	\section{Proposed Multi-Mode HAPS Communication Payload}
\begin{figure}
			\centering
		\includegraphics[width=0.9\linewidth]{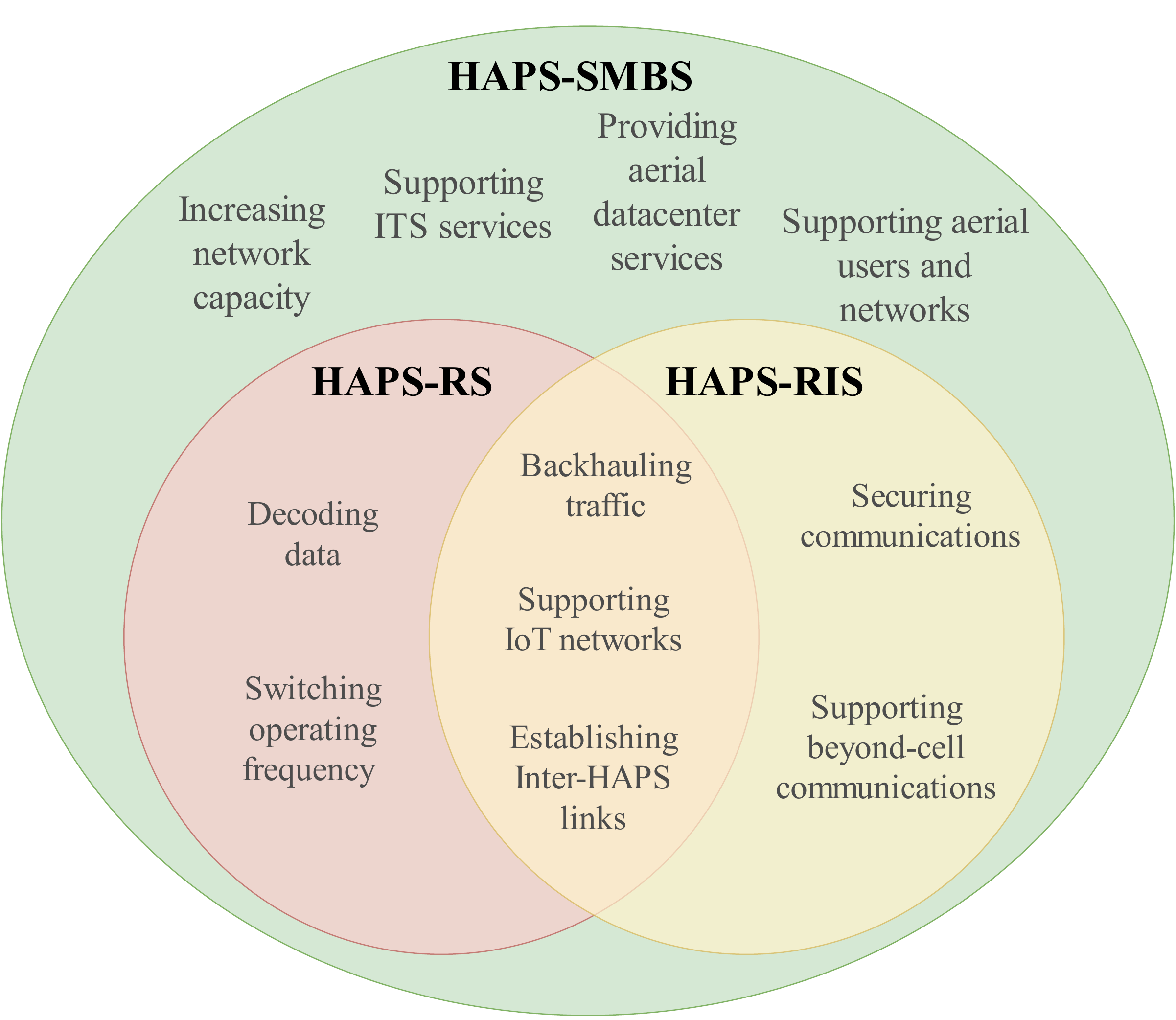}
		\caption{Relation between applications and enabling HAPS modes.}
	\label{fig:SMBS_RS_RIS_Apps}
	\end{figure}

	As discussed in the previous section, based on the HAPS payload type, different applications and use cases can be realized. The relation between a payload type and the suitable applications is shown in Fig. \ref{fig:SMBS_RS_RIS_Apps}.
	In general, as more active elements and power for processing are included in HAPS payload, more use cases can be served.
	However, this will inevitably increase  energy consumption and limit the mission duration. Also, demands may change dynamically, and so deploying a single-mode HAPS may not be cost-effective.
	
	To overcome the limits of a single mode payload, we propose a multi-mode HAPS payload, where both passive and active components can be integrated. For the passive payload, RIS layer can be integrated  on the underside of HAPS. Also, active components for signal processing and channel estimation can  be included. The active elements can be remotely controlled and switched ON/OFF opportunistically based on the use case demands and the application requirements. 
	The functioning mechanism of the multiple HAPS payload can be either based on a mode selection or cooperation-based approach. While a mode selection based approach maximizes the energy efficiency, a cooperation-based approach enhances reliability and communication performance.

    \subsection{HAPS Mode Selection} 

    \begin{figure}[t]
			\centering
		\includegraphics[width=1.12\linewidth]{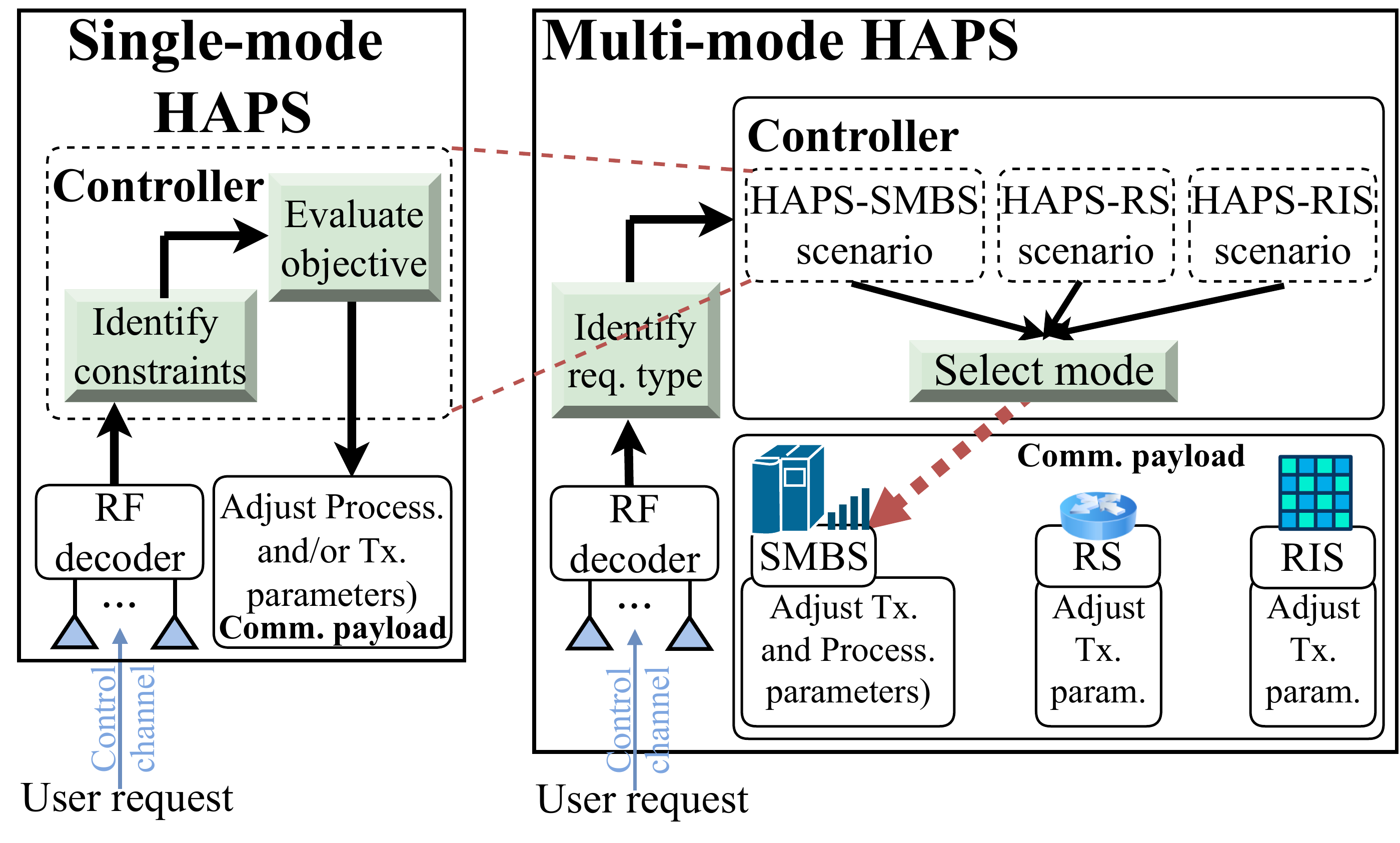}
		\caption{HAPS mode selection mechanism.}
		\label{fig:HAPS_modes}
	\end{figure}

    

  The mode selection mechanism is summarized in Fig. \ref{fig:HAPS_modes}. 
  We assume that the HAPS has a control circuit that exploits a low-power control channel to receive user requests. Whenever
a request is received, the HAPS starts  
identifying the nature of the request among different types, such as
\textit{communication}, \textit{content delivery}, \textit{caching} or \textit{task-offloading/computation}. This step is not required in a single-mode HAPS.

For a communication request, all modes compete with each other to maximize/minimize a predefined objective. For instance, assuming that the objective is to minimize the HAPS’s communication energy consumption in a backhauling scenario, then the HAPS-RIS mode might be selected if it satisfies the QoS requirements with minimal energy. If the objective is to maximize the backhauling capacity, then based on HAPS's location, available energy, and RIS size constraints, the best mode can be determined.

    When a content delivery request is received, the HAPS checks if it is able to satisfy it directly, i.e., if it has previously cached the content. If this is the case, then the HAPS-SMBS mode is activated and it precodes the content to send it to the user. Otherwise, HAPS-RS or HAPS-RIS mode is activated to forward the content from the core network to the user, via its ground communication gateway. If the pulled content is highly popular, HAPS-SMBS mode is switched on to proactively cache the content for future use.
    
    
    Finally, when a task offloading/computing request is received, the HAPS-SMBS mode evaluates its capacity to process the task and resulting energy consumption, given the task offloading latency constraint. In the meantime, both HAPS-RS and HAPS-RIS modes optimize their objectives if the task is forwarded to an available ground cloud/fog/edge server. The mode that is expected to achieve the best objective performance will be chosen and the related physical equipment parameters will be adjusted accordingly.

    \subsection{Benefits}
    A multi-mode HAPS presents several benefits. First, through its ability to adaptively switch between modes, it ensures that 
    the active components are used only for particular demands and use cases. Hence, passive components are often used, which significantly reduces energy consumption, and this ultimately extends the HAPS flight duration.
    Moreover, the availability of multiple modes guarantees robustness against equipment failure. For instance, if the active radio transceivers required for the HAPS-SMBS and HAPS-RS modes fail, the HAPS-RIS mode can take over temporarily, until the faulty equipment is fixed.
    Finally, performance can be enhanced through the combination of several modes. For instance, by using  RIS at the SMBS transmitter, reliable and high order modulation signals can be transmitted in an energy-efficient manner \cite{alfattani2021link}. Also, by simultaneously using the RIS and RS modes for signal forwarding and selection combining at the receiver, the communication performances in terms of capacity and outage probability are expected to improve, as demonstrated for RIS and RS equipped UAV systems  \cite{shafique2020optimization}.

	\subsection{Challenges and Research Directions}

    One of the most critical challenges to overcome in multi-mode HAPS is to respect the supported communication payload. Fortunately, several efforts are realized to reach this goal, including 5G gNBs miniaturization (below 50 kg), RIS weight reduction using lighter substrates, e.g., F4B \cite{Xilong2021}, and HAPS payload improvement with the deployment of large payload aircraft (dozens to hundreds of kilograms), e.g., the Elevate and Stratobus aircraft \cite{kurt2021vision}.
    
    Moreover, the design of an efficient mode-switching mechanism is a challenging task. The mechanism needs to consider  several parameters, e.g., user locations, channel state information (CSI), type of demand, current caching, and computing status of the HAPS and cloud server, which may not be all available in a timely manner for a rapid switching decision. Consequently, the mode selection strategy has to rely on partial or no prior knowledge of these parameters, and it must make decisions rapidly and be adaptable to the environment's traffic and mobility dynamics. To cope with these conditions, centralized or cooperative decentralized machine learning-based mode selection can be developed and evaluated, inspired by cloud radio access networks.

    Since computing and caching are supported by the HAPS-SMBS mode, problems of task offloading to the cloud and reactive/proactive caching at the HAPS must be rigorously treated. Although conventional solutions to task offloading and caching can be adopted, their development in the context of HAPS systems is still in its infancy \cite{Vallero2022}.

   Mode switching may seem straightforward when aiming to satisfy single or several homogeneous demands. However, operating a multi-mode selection to satisfy heterogeneous demands may become difficult and very complex since one mode may be adequate for a set of demands but not necessarily for a different set. Consequently, more advanced mode selection approaches should be developed while taking into account the heterogeneity of demands.
    To do so, game theory tools can be leveraged to evaluate the consequences of each mode and select the best strategy for modes selection. For highly dynamic environments, online learning methods, e.g., Q-learning, can be adopted for real-time mode switching. Furthermore, a combination of sophisticated modulation, e.g., orthogonal time-frequency space, of next-generation multiple access techniques like non-orthogonal/rate-splitting multiple access, and simultaneous activation of different modes, may be considered in such heterogeneous demand conditions.
    
     Finally, due to the limited energy onboard the multi-mode HAPS, advanced energy-saving techniques, such as simultaneous wireless
information and power transfer can be leveraged.
    

	\section{Case Study}
	In this case study, we discuss the differences between HAPS communication modes in terms of capacity, energy efficiency, and latency. Also, we identify the conditions favorable to the selection of one specific mode over another.
 
	\subsection{HAPS-RS versus HAPS-RIS}
\begin{figure}
		\centering
		\includegraphics[width=\linewidth]{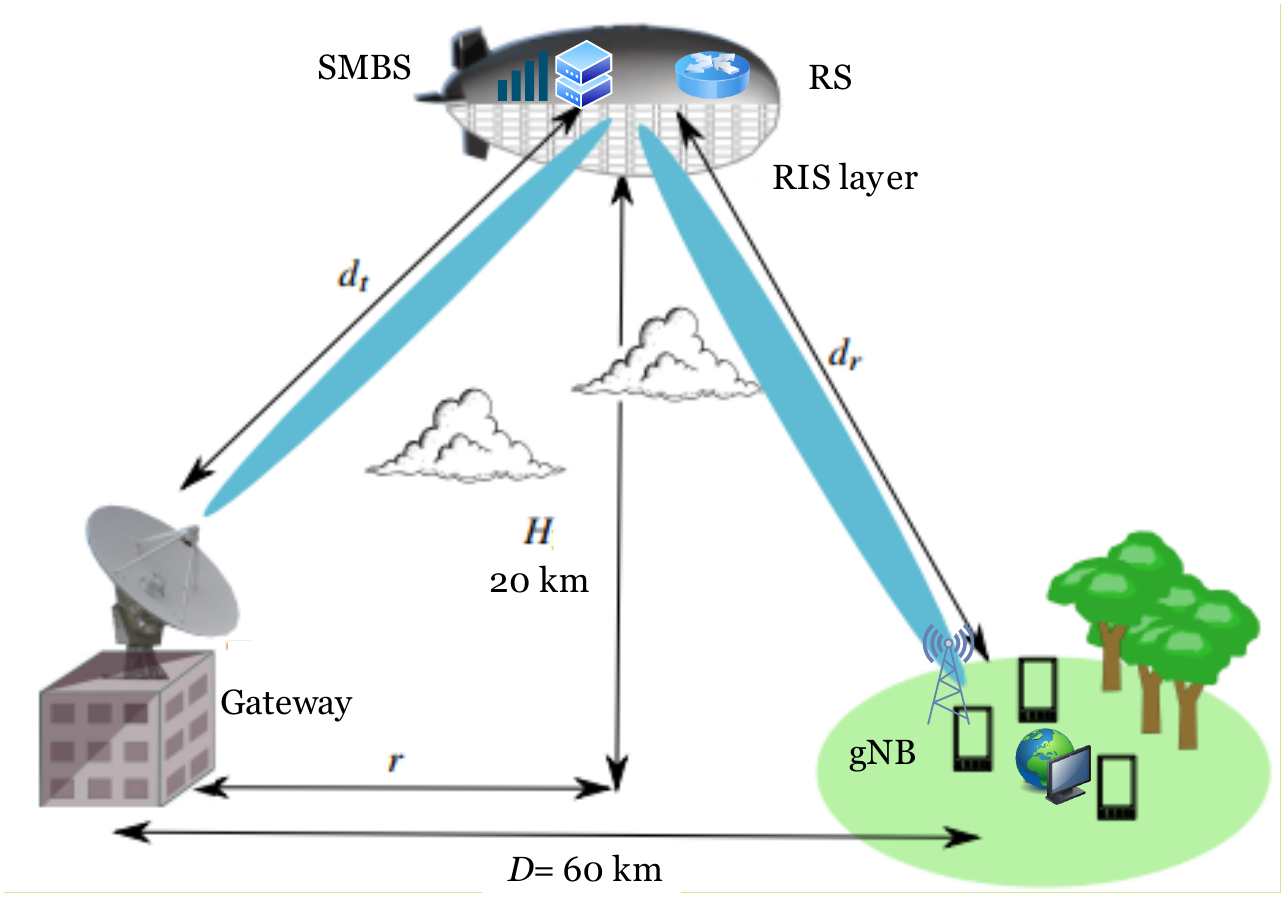}
	\caption{Backhauling system model using a multi-mode HAPS.}
		\label{fig:model}
	\end{figure}

	We consider a backhauling scenario for traffic from a gNB located at $D=60$ km away from its core network gateway, via a HAPS-RS mode or a HAPS-RIS mode, as depicted in Fig. \ref{fig:model}.
	We assume that the gNB is equipped with two antennas: The first antenna serves terrestrial users, while the second ensures communication with the HAPS. We assume that the gNB's transmit power and antenna gain are $P_{\rm gNB}=35$ dBm and $G_{\rm gNB}=15$ dB, respectively (TR 136.942).
	The gateway uses a highly directional antenna to communicate with the HAPS, having maximum power and gain set to $P_{0}^{\rm max} =33$ dBm and  $G_{0}^{\rm max} =43.2$ dB, respectively (TR 38.811). For the HAPS-RS, we consider a repetition-coded half-duplex decode-and-forward  relay  with antenna gain $G_{\rm RS}=15$ dB, while its payload power consumption is up to 1 kW, as specified for the X-Station HAPS developed by StratXX. Finally, for the HAPS-RIS communication payload, power consumption is issued from RIS phase-shifting, which depends on the used technology and phase-shift resolution. Typically, a 6-bit phase-shifting resolution	would consume around 7.8 mW per RIS reflecting element \cite{huang2019reconfigurable}. Thus, its total consumption is $N \times 7.8$ mW, where $N$ is the number of reflecting elements. 
	
For a fair comparison between the HAPS-RS and HAPS-RIS, we respect the power conservation principle. Specifically, the gateway's transmit power in the HAPS-RIS scenario ($P_0^{\rm max}$) is equal to the sum of transmit powers of the gateway ($P_{0}=\alpha P_0^{\rm max}$) and relay ($P_{RS}=(1-\alpha) P_0^{\rm max}$) in the HAPS-RS scenario, where $0<\alpha<1$. Finally, we set the operating frequency to 2 GHz\footnote{Although our case study is focused on radio frequency links, it can be easily extended to FSO systems operating at higher frequencies, by considering the relevant channel parameters.} in a rural environment with dry air atmospheric attenuation that has air pressure 101300 Pa, temperature 15$^\circ$C and   average tropospheric scintillation loss 0.5 dB. The details of the channel equations and capacity calculations for different environments are provided in \cite{alfattani2021link}.

\begin{figure}
		\centering
		\includegraphics[width=\linewidth]{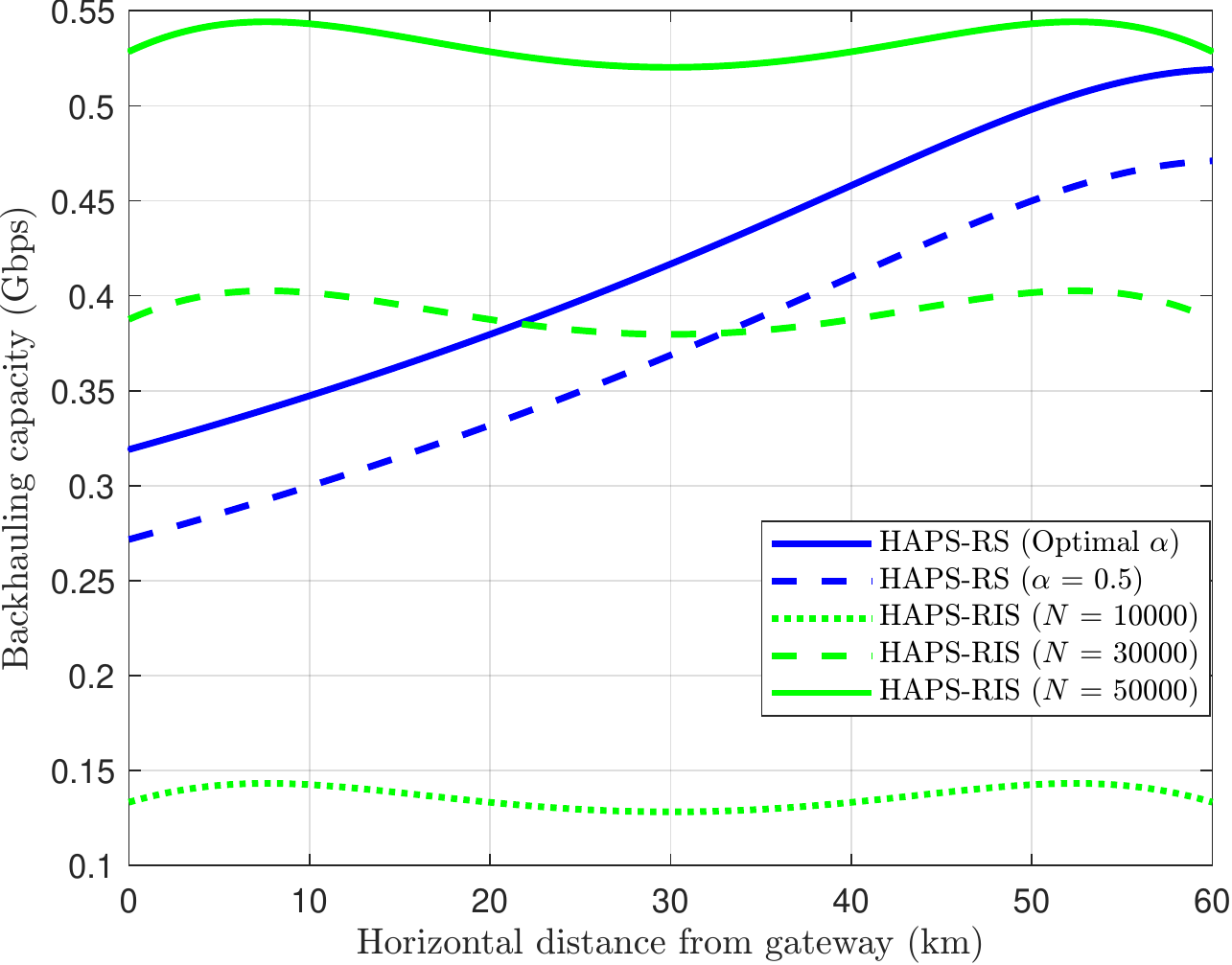}
		\caption{Backhauling capacity comparison between HAPS-RS and HAPS-RIS.  }
		\label{fig:RS_RIS_cap}
	\end{figure}
		\begin{figure}
	\centering
	\includegraphics[width=\linewidth]{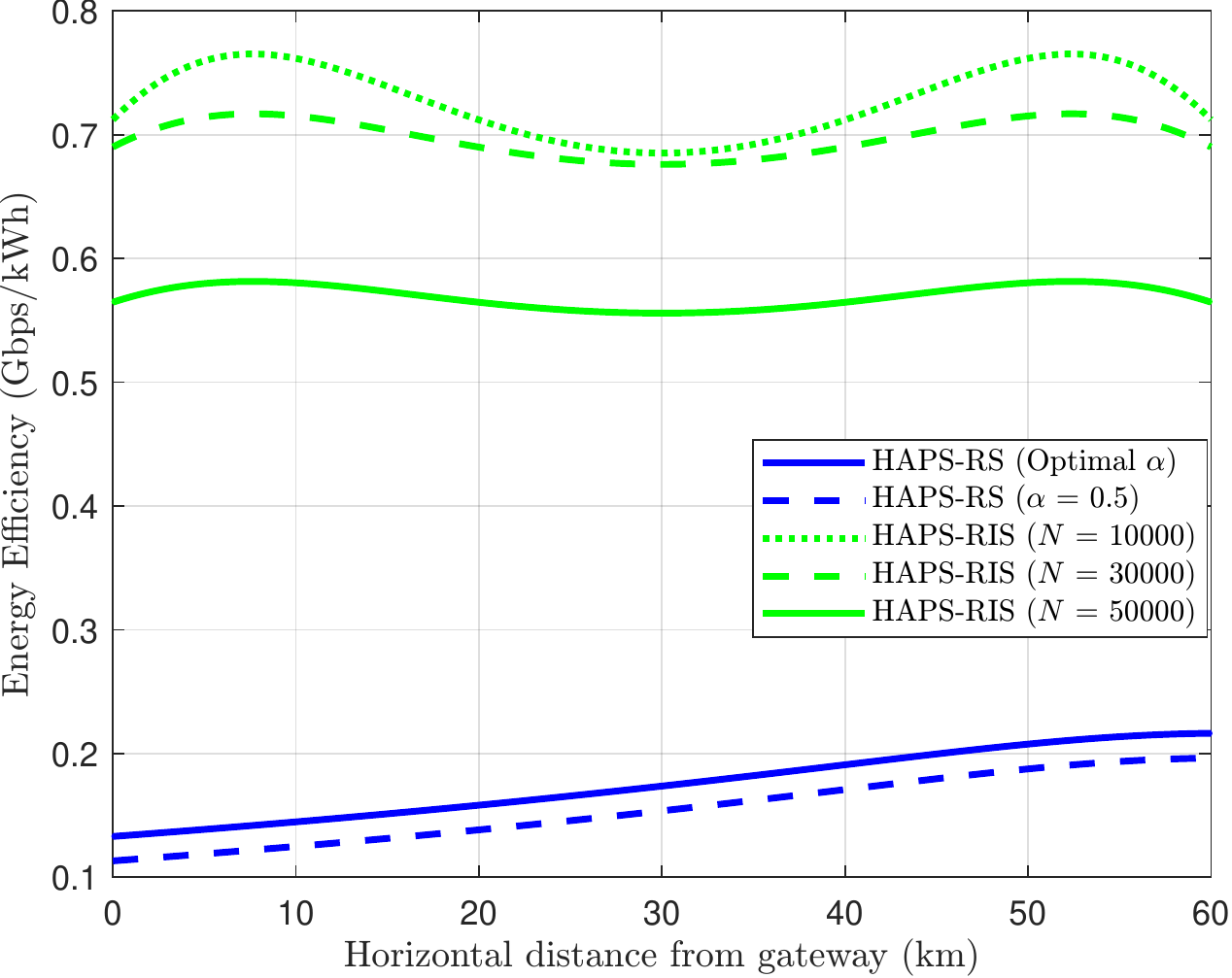}
	\caption{Energy efficiency  comparison between HAPS-RS and HAPS-RIS.}
	\label{fig:RS_RIS_energy_eff}
\end{figure}

Fig. \ref{fig:RS_RIS_cap} illustrates the backhauling capacity performance of the HAPS-RS and HAPS-RIS as a function of the horizontal distance between the HAPS and  gateway, where the horizontal distance is measured between the projection of the HAPS location on the ground and the gateway's location. 
For the HAPS-RS, we evaluated the capacity for fixed $\alpha=0.5$ and optimal $\alpha$ (obtained numerically), i.e., the one that achieves the best performance, while for the  HAPS-RIS, we evaluated the capacity for $N \in \{10000, 30000, 50000 \}$ \cite{Ellingson2021}.
As the distance from the gateway increases, i.e., HAPS gets closer to gNB, the HAPS-RS demonstrates a performance improvement with a maximal capacity achieved when it is on top of the gNB. Indeed, the decoding efficiency of the HAPS-RS improves when it gets closer to the gNB, while the gateway and its high antenna gains compensate for the degarded communication link between the HAPS-RS and the gateway. When $\alpha=0.5$, the capacity is degraded by 11\%, compared to the optimal case. 
Using the HAPS-RIS instead, we notice that two optimal locations are identified at distances of 6.9 km and 52.3 km. This result agrees with our findings in \cite{alfattani2021link}\footnote{According to \cite[Proposition 2]{alfattani2021link}, when $H<\frac{D}{2}$, the optimal HAPS-RIS locations are given by $\frac{D}{2} \pm \sqrt{\left( \frac{D}{2}\right)^2-H^2}$.}. Also, the backhauling capacity enhances with $N$. This is expected since a higher number of carefully configured reflecting elements $N$ allows the HAPS to generate a stronger reflection gain. 
From the backhauling capacity perspective, the mode selection between HAPS-RS and HAPS-RIS depends on several parameters, namely the HAPS's location and $N$. For instance, if $N=30000$, the HAPS-RS mode would be preferred when the horizontal distance from the gateway is above 21.5 km, while the HAPS-RIS mode would be recommended if the HAPS were closer to the gateway.


Fig. \ref{fig:RS_RIS_energy_eff} 	presents the related energy efficiency performance, defined as the ratio between the backhauling capacity and consumed energy. 
As we can see, the HAPS-RIS performs better for
any $N$ and location, 
due to the power consumption of the RIS being lower than that of active relaying.
Finally, as shown in Figs. \ref{fig:RS_RIS_cap}-- \ref{fig:RS_RIS_energy_eff}, HAPS-RIS mode demonstrates small performance variations with distance (around 5\%), which make it more resistant to location drifting than the HAPS-RS mode.

\vspace{-10pt}
\subsection{HAPS-SMBS versus HAPS-RS/HAPS-RIS}

\begin{figure}
	\centering
	\includegraphics[width=\linewidth]{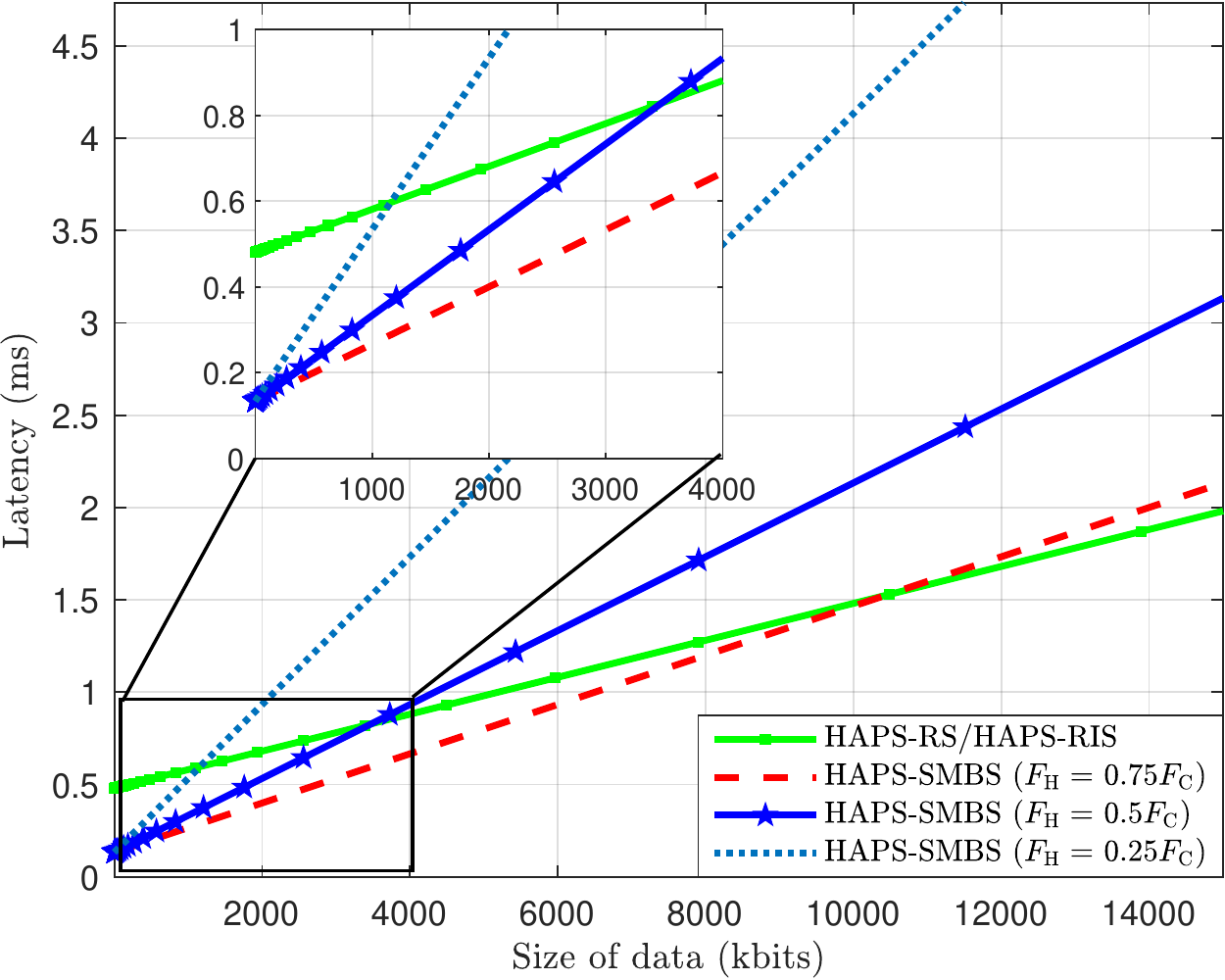}
	\caption{Latency comparison between HAPS-SMBS and HAPS-RS/HAPS-RIS.}
	\label{fig:latency_different_fH}
\end{figure}

We consider here a task-offloading scenario where a gNB can offload its task to the HAPS or to the cloud. Specifically, the HAPS-SMBS mode allows processing tasks using its onboard computation and caching capabilities, while the HAPS-RS/HAPS-RIS mode forwards the tasks to the gateway for processing within the cloud. Nevertheless, due to the relatively limited payload of the HAPS-SMBS, its computational capability $F_H$ (in CPU cycles/second) is typically lower than that of the cloud, denoted by $F_C$, while its communication latency is lower than that of the HAPS-RS/HAPS-RIS systems since the latter must reach the cloud via the ground gateway. The summation of the two, i.e., computation and communication latencies, defines the task-offloading latency. 

In Fig. \ref{fig:latency_different_fH}, we compare the performance in terms of task-offloading latency for the HAPS-SMBS and HAPS-RS/HAPS-RIS systems as a function of the task's data size, and given different values of $F_H$. For these simulations, we considered the same assumptions as in the previous subsection, except for the location of the HAPS, which is optimized according to the adopted mode\footnote{Optimal locations are chosen as follows: For HAPS-SMBS, it is set directly above the gNB; For HAPS-RS, it is determined numerically; For HAPS-RIS, it is calculated using  \cite[Proposition 2]{alfattani2021link}.}. 
Also, we assume that $F_{C} = 4 \times 10^9$ CPU cycles/second, and that each data unit requires $4$ CPU cycles to be processed.
Accordingly, we see that any mode's latency performance degrades linearly with the data size. 
Moreover, the HAPS-SMBS is the preferred mode for small data sizes, while HAPS-RS/HAPS-RIS modes are recommended for large data sizes and $F_H<F_C$. To conclude, the selection of the mode for task-offloading 
depends on several parameters, including the data size, HAPS location, $F_H$, and $N$.


  
\vspace{-8pt}  
  
	\section{Conclusion}	
In this article, we proposed a novel HAPS operating mechanism where the communication payload could switch between multiple modes, namely SMBS, RS, and RIS. First, we compared the HAPS features to those of other NTN systems, followed by a discussion about each HAPS mode, its characteristics, and potential applications. Then, we proposed our multi-mode HAPS vision, where the related mode selection mechanism, benefits, and potential challenges, were presented. Subsequently, we developed a case study to evaluate the performances of each HAPS mode and to identify the auspicious conditions for the selection of a specific mode among the available ones.

\vspace{-8pt}

	
	\bibliographystyle{IEEEtran}
	\bibliography{IEEEabrv,FINAL_VTM}

\begin{thebibliography}{10}
\providecommand{\url}[1]{#1}
\csname url@samestyle\endcsname
\providecommand{\newblock}{\relax}
\providecommand{\bibinfo}[2]{#2}
\providecommand{\BIBentrySTDinterwordspacing}{\spaceskip=0pt\relax}
\providecommand{\BIBentryALTinterwordstretchfactor}{4}
\providecommand{\BIBentryALTinterwordspacing}{\spaceskip=\fontdimen2\font plus
\BIBentryALTinterwordstretchfactor\fontdimen3\font minus
  \fontdimen4\font\relax}
\providecommand{\BIBforeignlanguage}[2]{{%
\expandafter\ifx\csname l@#1\endcsname\relax
\typeout{** WARNING: IEEEtran.bst: No hyphenation pattern has been}%
\typeout{** loaded for the language `#1'. Using the pattern for}%
\typeout{** the default language instead.}%
\else
\language=\csname l@#1\endcsname
\fi
#2}}
\providecommand{\BIBdecl}{\relax}
\BIBdecl

\bibitem{Geraci2022}
G.~Geraci, A.~Garcia-Rodriguez, M.~M. Azari, A.~Lozano, M.~Mezzavilla,
  S.~Chatzinotas, Y.~Chen, S.~Rangan, and M.~D. Renzo, ``What will the future
  of {UAV} cellular communications be? {A} flight from {5G} to {6G},''
  \emph{IEEE Commun. Surv. Tuts.}, vol.~24, no.~3, pp. 1304--1335, Third
  quarter 2022.

\bibitem{kurt2021vision}
G.~K. Kurt, M.~G. Khoshkholgh, S.~Alfattani, A.~Ibrahim, T.~S. Darwish, M.~S.
  Alam, H.~Yanikomeroglu, and A.~Yongacoglu, ``A vision and framework for the
  high altitude platform station ({HAPS}) networks of the future,'' \emph{IEEE
  Commun. Surveys Tuts.}, vol.~23, no.~2, pp. 729--779, Secondquarter 2021.

\bibitem{Wang2022}
R.~Wang, M.~A. Kishk, and M.-S. Alouini, ``Ultra-dense {LEO} satellite-based
  communication systems: A novel modeling technique,'' \emph{IEEE Commun.
  Mag.}, vol.~60, no.~4, pp. 25--31, Apr. 2022.

\bibitem{alam2021high}
M.~S. Alam, G.~K. Kurt, H.~Yanikomeroglu, P.~Zhu, and N.~D. {\DJ}{\`a}o, ``High
  altitude platform station based super macro base station constellations,''
  \emph{IEEE Commun. Mag.}, vol.~59, no.~1, pp. 103--109, Jan. 2021.

\bibitem{alfattani2021aerial}
S.~Alfattani, W.~Jaafar, Y.~Hmamouche, H.~Yanikomeroglu, A.~Yonga{\c{c}}oglu,
  N.~D. {\DJ}{\`a}o, and P.~Zhu, ``Aerial platforms with reconfigurable smart
  surfaces for {5G} and beyond,'' \emph{IEEE Commun. Mag.}, vol.~59, no.~1, pp.
  96--102, Jan. 2021.

\bibitem{alfattani2021link}
S.~Alfattani, W.~Jaafar, Y.~Hmamouche, H.~Yanikomeroglu, and
  A.~Yonga{\c{c}}oglu, ``Link budget analysis for reconfigurable smart surfaces
  in aerial platforms,'' \emph{IEEE Op. J. Commun. Soc.}, vol.~2, pp.
  1980--1995, 2021.

\bibitem{Cherif2021}
N.~Cherif, W.~Jaafar, H.~Yanikomeroglu, and A.~Yongacoglu, ``{3D} aerial
  highway: The key enabler of the retail industry transformation,'' \emph{IEEE
  Commun. Mag.}, vol.~59, no.~9, pp. 65--71, Sep. 2021.

\bibitem{jaafar2022haps}
W.~Jaafar and H.~Yanikomeroglu, ``{HAPS}-{ITS}: Enabling future {ITS} services
  in trans-continental highways,'' \emph{IEEE Commun. Mag.}, vol.~60, no.~10,
  pp. 80--86, Oct. 2022.

\bibitem{long2020reflections}
H.~Long, M.~Chen, Z.~Yang, B.~Wang, Z.~Li, X.~Yun, and M.~Shikh-Bahaei,
  ``Reflections in the sky: Joint trajectory and passive beamforming design for
  secure {UAV} networks with reconfigurable intelligent surface,'' \emph{arXiv
  preprint arXiv:2005.10559}, 2020.

\bibitem{alfattani2022beyond}
S.~Alfattani, A.~Yadav, H.~Yanikomeroglu, and A.~Yongacoglu, ``Beyond-cell
  communications via {HAPS}-{RIS},'' in \emph{Proc. IEEE Glob. Commun. Wkshp.
  (GLOBECOM)}, 2022, pp. 1--6.

\bibitem{shafique2020optimization}
T.~Shafique, H.~Tabassum, and E.~Hossain, ``Optimization of wireless relaying
  with flexible {UAV}-borne reflecting surfaces,'' \emph{IEEE Trans. Commun.},
  vol.~69, no.~1, pp. 309--325, Jan. 2021.

\bibitem{Xilong2021}
X.~Pei, H.~Yin, L.~Tan, L.~Cao, Z.~Li, K.~Wang, K.~Zhang, and E.~Björnson,
  ``{RIS}-aided wireless communications: Prototyping, adaptive beamforming, and
  indoor/outdoor field trials,'' \emph{IEEE Trans. Commun.}, vol.~69, no.~12,
  pp. 8627--8640, Dec. 2021.

\bibitem{Vallero2022}
G.~Vallero, D.~Renga, and M.~Meo, ``Caching in the air: High altitude platform
  stations for urban environments,'' in \emph{Proc. IEEE Wireless Commun.
  Network. Conf. (WCNC)}, 2022, pp. 2244--2249.

\bibitem{huang2019reconfigurable}
C.~Huang, A.~Zappone, G.~C. Alexandropoulos, M.~Debbah, and C.~Yuen,
  ``Reconfigurable intelligent surfaces for energy efficiency in wireless
  communication,'' \emph{IEEE Trans. Wireless Commun.}, vol.~18, no.~8, pp.
  4157--4170, Aug. 2019.

\bibitem{Ellingson2021}
S.~W. Ellingson, ``Path loss in reconfigurable intelligent surface-enabled
  channels,'' in \emph{Proc. IEEE Ann. Int. Symp. Perso. Indoor Mob. Radio
  Commun. (PIMRC)}, 2021, pp. 829--835.

\end{thebibliography}

\vspace{-8pt}
\section*{Biographies}
\small{
\noindent \textbf{Safwan Alfattani} [M] (smalfattani@kau.edu.sa) 
is an Assistant Professor at King AbdulAziz University, Saudi Arabia. He received his M.Sc. degree 
and Ph.D from  University of Ottawa, Canada. In 2022, He was the recipient of the grand prize of the IEEE Future Networks competition on NTN for B5G and 6G.
His research interests include wireless communications, NTN, RIS, and IoT networks.
\\
\\
\noindent \textbf{Wael Jaafar} [SM] (wael.jaafar@etsmtl.ca) is a Professor at ÉTS Montreal, Canada. His research interests include wireless communications, edge computing, and machine learning. He 
received
prestigious grants such as NSERC Alexandre Graham-Bell (BESC) and FRQNT internship scholarship.
\\
\\
\noindent \textbf{Halim Yanikomeroglu} [F] (halim@sce.carleton.ca)  is a  Professor at Carleton University, Canada. His research interests
cover many aspects of 6G wireless networks. His research with industry has resulted in 39 granted patents. He is a Fellow of IEEE, the
Engineering Institute of Canada, and the Canadian Academy of Engineering, and he is a Distinguished Speaker for
the IEEE Communications Society and IEEE Vehicular
Technology Society.
\\
\\
\noindent \textbf{Abbas Yongaçoglu} [LM] (yongac@uottawa.ca) is an Emeritus Professor at the University of Ottawa, Canada.
His area of research is wireless communications and signal processing.

}

\end{document}